\documentclass[11pt,a4paper]{article}

\usepackage[margin=1in]{geometry} % Standard 1-inch margins
\usepackage{amsmath,amssymb,amsthm}
\usepackage{graphicx}
\usepackage{bm}
\usepackage[hidelinks]{hyperref}
\usepackage{authblk} % Allows your existing \author and \affil blocks to work natively

\begin{document}

\title{Euler-Korteweg vortices: A fluid-mechanical analogue to the Schr\"odinger and Klein–Gordon equations}

%%%% To generate auto affiliation numbers please use \author{}\affil{} command

\author{Daniël Bischoff van Heemskerck}
\affil{d.m.f.bischoff@nescio.foundation}
\date{\small December 29, 2025 (Revised: June 4, 2026)}
%\author{Insert third author name here}
%\author[3]{Insert fourth author name here} %%% Use optional bracket [3] to change the respective address
%\affil{Insert third author address here}

%\author{Insert last author name here\thanks{These authors contributed equally to this work}}
%\affil{Insert last author address here}

%%% To include the collaborator name... Please use the command "\collaborator"
%%% For example: \collaborator{ATLAS Collaboration}
\maketitle

\begin{abstract}%
Quantum theory and relativity exhibit several formal analogies with fluid mechanics. This paper extends upon known analogies by showing that under specific assumptions, a classical ideal fluid can be cast into equations that are mathematically equivalent to the Schr\"odinger and Klein–Gordon equations. By assuming that the angular momentum of an irrotational vortex in an inviscid, barotropic, isothermal fluid with sound speed $c$ is equal in magnitude to the reduced Planck constant, and incorporating Korteweg capillary stress, a complex wave equation describing the momentum and continuity equations of an Euler–Korteweg vortex is obtained. When uniform convection is introduced, the weak field approximation of this wave equation is formally equivalent to Schr\"odinger’s equation. The model is shown to yield analogues to de Broglie wavelength, the Einstein–Planck relation, the Born rule and the uncertainty principle. Accounting for the retarded propagation of the wave field of a vortex in convection requires the Lorentz transformation and yields a wave equation mathematically equivalent to the Klein–Gordon equation, with Schr\"odinger’s equation appearing as the low-Mach-number limit. 
\end{abstract}

\section{Introduction}

The analogy between relativity and fluid models is well known \cite{barcelo2005}. Unruh \cite{unruh1976,unruh1981,unruh1995} and Visser \cite{visser1993,visser1998} demonstrated that acoustic perturbations in an inviscid, barotropic fluid obey a Lorentzian spacetime, which becomes curved in convection, deriving the corresponding acoustic metric complete with event horizons. Laboratory experiments by Weinfurtner et al. \cite{weinfurtner2011} verified this correspondence experimentally by observing Hawking-like emission in sink flow. Jacobson showed the analogy extends beyond continuum mechanics \cite{jacobson1995}, recovering the Einstein field equations as the equation of state of a thermodynamic system.

Quantum mechanics also bears many correspondences with fluid mechanics. De Broglie \cite{debroglie1923} proposed that not only light but matter also can be described as a wave, and Schr\"odinger extended upon this idea, formulating a probabilistic wave equation describing fundamental particles \cite{schrodinger1926a,schrodinger1926b}. Madelung showed that this archetypal quantum equation can be written in hydrodynamic form as a complex function of a continuity and Euler equation \cite{madelung1927}, demonstrating a deep formal resemblance between quantum and fluid mechanics, which was later expanded by Takabayasi \cite{takabayasi1952}. Bohm, postulating particles to possess a quantum potential, developed de Broglie’s proposal into a causal interpretation in which particles follow determined trajectories guided by a guiding wave, yet exhibit probabilistic quantum behavior \cite{bohm1952a,bohm1952b}. This notion finds a striking correspondence in macroscopic hydrodynamic quantum analogs, where droplets walking on a vibrating fluid surface are guided by self-generated waves and display behavior such as interference and quantized orbits \cite{couder2005,couder2006,bush2015}. Furthermore, studies and experiments with superfluids, such as Bose-Einstein condensates (BECs), which are upon approximation inviscid, barotropic fluids, have shown that vortices in continuous fluids can display quantized behavior \cite{onsager1949,feynman1955,fetter2001}, and Popov has shown BEC vortices behave analogously to charged particles \cite{popov1973}. More recent studies have shown that a Korteweg-Euler system is formally analogous to the Gross Pitaevskii non-linear Schr\"odinger equation if the capillary coefficient is \(K = 1/(4\rho)\) \cite{benzoni2005,carles2012,bresch2019,mauri2021,gui2025}. That is, quantum or Bohm potential is formally analogous to Korteweg stress, a capillary force caused by a steep free energy gradient, proposed by van der Waals \cite{vanderwaals1893} for classical fluids, and developed by Korteweg \cite{korteweg1901}, Dunn and Serrin \cite{dunnserrin1985}. Also, Vadasz has proposed that quantum behavior may arise from classical fluid dynamics if subatomic particles are treated as compressible fluids.\cite{10.1063/5.0271672}

Similar to analogue gravity, where fluid conditions are explicitly chosen to recover Lorentzian spacetimes, this paper investigates the exact theoretical conditions under which a classical continuum mathematically recovers quantum wave equations. 

The manuscript extends upon previous work by showing that beyond established correspondences, the mathematical structures of the Schr\"odinger and Klein-Gordon equations can be derived entirely as the low-Mach and high-Mach limits of a strictly classical, 'Newtonian' continuum model, provided highly specific assumptions are met.
 
Whereas Madelung's formulation and Bohmian mechanics cast quantum equations
into hydrodynamic form, and Gross--Pitaevskii theory describes superfluids in
which quantized circulation follows from the single-valuedness of a macroscopic
quantum wavefunction, the novelty of the present work lies in applying a reverse Madelung transform to
a classical Euler--Korteweg vortex. In this model, properties such as a
single-valued phase and quantized circulation are not introduced as quantum
postulates, but arise from the classical topology of an irrotational vortex.

By considering a model for an inviscid, barotropic, isothermal fluid with \(c_s = c\), assuming this fluid sustains vortices that (1) are irrotational, (2) have an angular momentum equal to \(\hbar\) and (3) contain a steep pressure gradient at their core with a certain ratio between characteristic lengths, causing capillary Korteweg-stress with coefficient \(K = \hbar^2/(4m^2 \rho)\), a set of structural assumptions is identified that yields a mathematically consistent set of analogies between fluid dynamics and the formalism of certain quantum and relativistic wave equations.
 
 Although analogies can be of theoretical interest, it must be noted that the present work does not claim to provide a physical reconstruction of quantum mechanics or relativity.

%%%%%%%%%%%%%%%%%%%%%%%%%%%%%%%%%%%%%%%%%%

\section{The equation of state}

Let us consider an inviscid, isothermal and
barotropic fluid, where
\begin{equation}
c_s^2 = \frac{dP}{d\rho} = \frac{1}{\rho \beta} = \text{constant,}
\label{eq:EOS1}
\end{equation}

where \(P\) and \(\rho\) are respectively the pressure and density of the
fluid, and \(\beta\) its compressibility. For the purpose of retrieving the analogous Schr\"odinger equation, we shall set \(c_s = c\) with \(c\) being the speed of light.  
And so the equation of state of this fluid is:
\begin{equation}
P = \rho c^2
\label{eq:EOS2}
\end{equation}

Considering the value of $c$ it is clear the product $\rho \beta$ is so small that such an equation of state is purely hypothetical and does not represent any classical known fluid.

\section{Quantized circulation and the Planck constant}

Let us imagine an irrotational vortex as an excitation of the fluid. For an
irrotational flow the velocity field in polar coordinates is given by the
gradient of a potential:
\begin{equation}
v(r,\theta) = \nabla \Phi(r,\theta)
\label{eq:2.1}
\end{equation}

\begin{equation}
v_r = \frac{\partial \Phi}{\partial r}, \qquad
v_\theta = \frac{1}{r}\frac{\partial \Phi}{\partial \theta}
\label{eq:2.2}
\end{equation}

\begin{equation}
\nabla\Phi = 
\frac{\partial \Phi}{\partial r}\,\hat{r}
+ \frac{1}{r}\frac{\partial \Phi}{\partial \theta}\,\hat{\theta}
\label{eq:2.3}
\end{equation}

And we establish any radial flow as axisymmetric, being purely dependent on
\(r\) and not \(\theta\):
\begin{equation}
\frac{\partial v_r}{\partial \theta} = 0
\label{eq:2.4}
\end{equation}

And thus, the vorticity satisfies:
\begin{equation}
\nabla \times v 
= \frac{1}{r}\frac{\partial}{\partial r}(r v_\theta)
 - \frac{1}{r}\frac{\partial v_r}{\partial \theta}
= \frac{1}{r}\frac{\partial}{\partial r}(r v_\theta)
= 0
\label{eq:2.5}
\end{equation}

Which ensures the vortex is irrotational, even though it can support a
radial flow, and ensures the circulation of the system \(\Gamma\) and the
mass-specific angular momentum $\ell$ are:
\begin{equation}
\Gamma = 2\pi v_\theta r = \text{constant}
\label{eq:2.6}
\end{equation}

\begin{equation}
\ell = v_\theta r = \text{constant}
\label{eq:2.7}
\end{equation}

For all \(r\), as the velocity of the vortex diminishes in exact
conjunction with increasing \(r\).

\begin{equation}
\ell = \frac{\partial \Phi}{\partial \theta}
\label{eq:2.8}
\end{equation}

And integrating \(\nabla \Phi\) with respect to \(\theta\), since
\(\partial v_r / \partial \theta = 0\), we find:
\begin{equation}
\Phi(\theta) = \ell\,\theta
\label{eq:2.9}
\end{equation}

And:
\begin{equation}
\nabla\Phi(\theta) = \ell\,\nabla \theta
\label{eq:2.10}
\end{equation}

Any contour of the circulation \(\Gamma\) is the line integral of the
velocity around a closed loop enclosing the vortex:
\begin{equation}
\Gamma = \int v \cdot dl
= \ell \int \nabla\theta \cdot dl
= \ell\,\Delta^{t}\theta
\label{eq:2.11}
\end{equation}

Where \(\Delta^t \theta\) is the total coordinate distance. Because the
coordinate \(\theta\) must return to its initial value after an integer
number of turns, modulo \(2\pi\), around the vortex, the circulation is
essentially quantized:
\begin{equation}
\Delta^t \theta = 2\pi n
\label{eq:2.12}
\end{equation}

Where \(n \in \mathbb{Z}\) is the integer winding number. Thus, we have:
\begin{equation}
\Gamma = \int v\cdot dl = 2\pi n \ell
\label{eq:2.13}
\end{equation}

The simplest is the case where \(n=1\), which we shall discuss at present.

First let us ascribe an effective mass \(m\) to the vortex. While this is
not a conventional procedure in classical fluid mechanics, it is well
established in theory for superfluid vortices.\cite{feynman1955,thouless2007,Richaud2020,Richaud2021,kanjo2024,levrouw2025,Richaud2025,PhysRevE.111.034216}

With this, we ascribe a total angular momentum \(J\) to the vortex:
\begin{equation}
J = m \ell
\label{eq:2.14}
\end{equation}

Which is constant if \(m\) is. For the sake of clarity, we will henceforth assume:
\begin{equation}
J = \hbar, \qquad h = 2\pi\hbar
\label{eq:2.15}
\end{equation}

Where \(\hbar\) is the reduced Planck constant and \(h\) Planck’s constant.

While the dimensions and constancy of $J$ are physically motivated by classical vortex dynamics, its specific value is a matter of parameterization. The following mathematical structure holds for any arbitrary constant $J$ and sound speed $c_s$. We explicitly assign $J = \hbar$ and $c_s = c$ here solely to exhibit the conditional equivalence to the Schr\"odinger and Klein-Gordon equations.

Also note that this is the same winding-number mechanism that underlies the usual circulation quantization of superfluid vortices, although here vortex topology enforces an effective quantization of circulation, rather than being introduced through the single-valuedness of a complex order parameter.

\section{Korteweg stress as quantum potential}\label{sec3}

Thus, we can already in this manner see that mathematically, a classical vortex can exhibit
‘quantized’ behavior akin to quantum systems. Furthermore, the periodicity of
$\theta$ does allow us to define it as a single valued phase, and so we can formulate
a complex field or wavefunction $\psi$ describing the fluid excitation through
both the density and the phase:
\begin{equation}
\psi = \sqrt{\rho}\, e^{i\theta}
\label{eq:3.1}
\end{equation}

Where $\rho$ is the mass density, $\theta = \Phi / (\hbar/m)$ is the azimuthal
angle and the phase, $e$ Euler’s number and $i$ the imaginary unit.
Inserting $\hbar = m\ell$ into eq.~\ref{eq:2.10} we have:
\begin{equation}
v = \nabla\Phi = \frac{\hbar}{m} \nabla\theta
\label{eq:3.2}
\end{equation}

Which is formally the same expression for a particle’s translational velocity
or the ‘guiding wave equation’ in Bohm’s
pilot wave theory~\cite{bohm1952a,bohm1952b},
if we interpret the azimuthal coordinate $\theta$ as the phase of the wavefunction.

Remembering the continuity equation:
\begin{equation}
\frac{\partial \rho}{\partial t} + \nabla \cdot (\rho v) = 0
\label{eq:3.3}
\end{equation}

And substituting eq.~\ref{eq:3.2}, we find something similar to the quantum
probability current equation~\cite{madelung1927}:
\begin{equation}
\frac{\partial \rho}{\partial t} + \frac{\hbar}{m}\,\nabla\cdot(\rho\nabla\theta) = 0
\label{eq:3.4}
\end{equation}

Although the correspondence is formal, still, $\rho$ is mass density, not probability density.

Euler's equation without external forces is:
\begin{equation}
\frac{\partial v}{\partial t} + (v\cdot\nabla)v = -\nabla h_{\rho}
\label{eq:3.5}
\end{equation}

Where the enthalpy is:
\begin{equation}
\nabla h_{\rho} = \frac{1}{\rho}\,\nabla P
\label{eq:3.6}
\end{equation}

But, if the vortex has a steep pressure gradient, implying a density defect
such as a spherical mass cavity or spike in the centre, we should account for
the free energy gradient $f_0$:
\begin{equation}
F = \int \left( f_0 + \frac{K(\rho)}{2} |\nabla\rho|^2 \right) dV,
\qquad K(\rho) = \frac{\kappa}{\rho}
\label{eq:3.7}
\end{equation}

Where $K$ is the capillary coefficient and $\kappa$ a constant to be determined.
The Euler–Korteweg equation, accounting for capillary Korteweg stress, and any
possible external force per unit mass is:
\begin{equation}
\frac{\partial v}{\partial t} + (v\cdot\nabla)v
= -\frac{1}{\rho}\nabla p + \frac{1}{\rho}\nabla\cdot\sigma^K + f_{\mathrm{ext}}
\label{eq:3.8}
\end{equation}

Where the Korteweg stress per unit mass expressed via $\kappa$ is:
\begin{equation}
\frac{1}{\rho}\,\nabla\cdot\sigma^K
= 2\kappa \,\nabla\!\left( \frac{\nabla^2 \sqrt{\rho}}{\sqrt{\rho}} \right)
\label{eq:3.9}
\end{equation}

And the arbitrary external force per unit mass due to an external scalar
potential $V$ is $f_{\mathrm{ext}} = -\frac{1}{m}\nabla V$.

Using the identity:
\begin{equation}
\frac{\nabla^2 \sqrt{\rho}}{\sqrt{\rho}}
= \frac{1}{2\rho}\nabla^2\rho - \frac{1}{4\rho^2}|\nabla\rho|^2
\label{eq:3.10}
\end{equation}

It follows that:
\begin{equation}
\frac{\kappa}{\rho}\nabla^2\rho
 - \frac{\kappa}{2\rho^2}|\nabla\rho|^2
= 2\kappa\,\frac{\nabla^2\sqrt{\rho}}{\sqrt{\rho}}
\label{eq:3.11}
\end{equation}

And if we set:
\begin{equation}
\kappa = \frac{\hbar^2}{4m^2}
\label{eq:3.12}
\end{equation}

Then we see the form of the Bohm quantum potential $Q$, which
Madelung~\cite{madelung1927} employed:
\begin{equation}
Q = -\frac{\hbar^2}{2m}\,\frac{\nabla^2\sqrt{\rho}}{\sqrt{\rho}}
\label{eq:3.13}
\end{equation}

\begin{equation}
\frac{1}{\rho}\,\nabla\cdot\sigma^K
= \frac{\hbar^2}{2m^2}\,\nabla\!\left(
\frac{\nabla^2\sqrt{\rho}}{\sqrt{\rho}}
\right)
= -\frac{1}{m}\nabla Q
\label{eq:3.14}
\end{equation}

Now setting $\kappa$ as such seems at first a quantum insertion, but it is good to note that we are working with the specialized assumption $J=\hbar$, and that $J^2/m^2=\ell^2$.

Following square gradient or mean field theory%
~\cite{cahn1958,cahn1959,rowlinson1982,sengers1991,chaikin1995,onuki2002},
the susceptibility or response function $\chi$ near critical temperature of a fluid interface $T_c$ is:
\begin{equation}
\chi = \left( \frac{\partial \rho}{\partial \mu} \right)_T
\label{eq:3.15}
\end{equation}

Where $\mu$ is the chemical potential and:
\begin{equation}
a = a_0 (T - T_c)
\label{eq:3.16}
\end{equation}

Then:
\begin{equation}
\chi =
\begin{cases}
1/a, & T > T_c \\
1/(2|a|), & T < T_c
\end{cases}
\label{eq:3.17}
\end{equation}

The characteristic correlation length $\xi$ is:
\begin{equation}
\xi =
\begin{cases}
\sqrt{K/a}, & T > T_c \\
\sqrt{K / (2|a|)}, & T < T_c
\end{cases}
\label{eq:3.18}
\end{equation}

Following the isothermal assumption:
\begin{equation}
\left( \frac{\partial\mu}{\partial p} \right)_T = \frac{1}{\rho},
\qquad
a = \left( \frac{\partial\mu}{\partial\rho} \right)_T
= \left( \frac{\partial\mu}{\partial p} \right)_T
  \left( \frac{\partial p}{\partial\rho} \right)_T
= \frac{c^2}{\rho}
\label{eq:3.19}
\end{equation}

Define a characteristic radius $r_c$ where the vortex azimuthal velocity is $c$:
\begin{equation}
r_c = \frac{\hbar}{mc},
\qquad
\frac{\hbar^2}{4m^2} = \frac{c^2 r_c^2}{4}
\label{eq:3.20}
\end{equation}

If:
\begin{equation}
\xi = \frac{r_c}{\sqrt{q}}
\label{eq:3.21}
\end{equation}

Then:
\begin{equation}
\kappa = c^2 \xi^2 = \frac{\hbar^2}{4m^2}
\label{eq:3.22}
\end{equation}

For $q = 4$ ($\xi_{+}$) or $q = 8$ ($\xi_{-}$) for $T > T_c$ and $T < T_c$ respectively.

Thus, Korteweg stress can take the form of Bohm potential under the specialized choice of $\kappa = \hbar^{2} / 4m^{2}$. Since $\hbar$ is here the specialized constant angular momentum of the system, and $\hbar^2/m^2 = \ell^2$, the specific form of $\kappa$ required for this exact correspondence does not need to be arbitrarily imposed. As shown here, it can be derived from square gradient mean-field theory under an assumption regarding a ratio between characteristic lengths. This shows, at least, that there are possibly physical solutions to the setting of $\kappa$.

As a note of interest, the correlation length $\xi$ is close to the healing length
$\xi_h$ in BEC vortices, and $\xi_h = \sqrt{\hbar^{2} / 2m^{2} c^{2}}$ if
$gn = mc^{2}$. Also, $\xi_{+}/\xi_{-} = 2$ is analogous to the ratio between the
characteristic coherence length for the superconductive phase $T < T_c$ and phase
$T > T_c$ in Ginzburg--Landau theory for superconductivity. If one identifies
$m^{*} c^{2} = |\alpha_{\mathrm{GL}}|$, the Ginzburg--Landau coherence lengths for
$\xi_{+}$ and $\xi_{-}$ are then described by eq.~\ref{eq:3.21} with $q = 4$ and
$q = 2$ respectively. 

\section{Schr\"odinger's equation}

Continuing, the full integrated Euler--Korteweg reads:
\begin{equation}
\frac{\partial \Phi}{\partial t}
+ \frac{1}{2} (\nabla \Phi)^2
- \frac{\hbar^{2}}{2 m^{2}}
  \frac{\nabla^{2} \sqrt{\rho}}{\sqrt{\rho}}
+ \frac{1}{m} V
= -h_{\rho}
\label{eq:4.1}
\end{equation}

Let us consider an observer at rest with respect to a set of
$x, y, z$ coordinates and the far field fluid. The fluid has a
constant drift velocity $v_{d}$ relative to the vortex, which is thus
seen to be in motion relative to the observer $(x,y,z)$ with $v_{d}$.

In the vortex frame we then have the total velocity potential being
equal to the rest velocity potential plus the drift potential:
\begin{equation}
\Phi(x,y,t) = \Phi_{c} + \Phi_{d}
\label{eq:4.2}
\end{equation}

With:
\begin{align}
x &= r\cos\theta, &
y &= r\sin\theta, \\
\Phi_{c} &= \frac{\hbar}{m}\,\theta(x,y)
          = \frac{\hbar}{m}
            \arctan\!\left(\frac{y}{x}\right)
\label{eq:4.3}
\end{align}

Since we have
\begin{align}
\nabla \Phi_{d} &= v_{d} = v_{d}\,\hat{x}, \\
\frac{\partial \Phi_{d}}{\partial x} &= v_{d}, &
\frac{\partial \Phi_{d}}{\partial y} &= 0, \\
\frac{\partial \Phi_{d}}{\partial t} &=
-\frac{1}{2}\left|\nabla \Phi_{d}\right|^{2}
\label{eq:4.4}
\end{align}

We can express:
\begin{equation}
\Phi_{d} = v_{d} x - \frac{1}{2} v_{d}^{2} t
\label{eq:4.5}
\end{equation}

In the drift case, the added phase becomes:
\begin{equation}
\theta_{d} = \frac{m v_{d}}{\hbar}\, x
- \frac{1}{2} \frac{m v_{d}^{2}}{\hbar} t
\label{eq:4.6}
\end{equation}

Which added to the `rest' phase becomes the total phase:
\begin{equation}
\theta = \theta_{c}
+ \frac{m v_{d}}{\hbar}\, x
+ \left( -\frac{1}{2} \frac{m v_{d}^{2}}{\hbar} t \right)
\label{eq:4.7}
\end{equation}

Now, to formulate the complex vortex wave equation, recall that:
\begin{equation}
\psi = \sqrt{\rho}\, e^{i\theta},
\qquad
\theta = \frac{m\Phi}{\hbar}
\label{eq:4.8}
\end{equation}

For the Schr\"odinger equation to appear we must neglect the azimuthal
flow field, which could be justified in a far field approximation where:
\begin{equation}
\Phi \approx \Phi_{d}
\label{eq:4.10}
\end{equation}

And ansatz $h_{\rho} = 0$, which can be justified when the enthalpy is
approximately constant, such as in the far field and low Mach-limit regime.

Then in the same manner as Madelung has shown, we can insert the Euler
equation and the continuity equation into eq.~\ref{eq:4.8}:
\begin{equation}
\psi = \sqrt{\rho}\, e^{i\theta}
\qquad
\begin{cases}
\displaystyle
\frac{\partial \Phi}{\partial t}
+ \frac{1}{2}(\nabla\Phi)^{2}
- \frac{\hbar^{2}}{2m^{2}}
  \frac{\nabla^{2}\sqrt{\rho}}{\sqrt{\rho}}
+ \frac{1}{m}V = 0
\\[0.8em]
\displaystyle
\frac{\partial \rho}{\partial t}
+ \nabla \cdot (\rho v) = 0
\end{cases}
\label{eq:4.11}
\end{equation}

And find:
\begin{equation}
i\hbar\,\frac{\partial \psi}{\partial t}
=
\left[
-\frac{\hbar^{2}}{2m}\nabla^{2}
+ V
\right]
\psi
\label{eq:4.12}
\end{equation}

Which formally resembles the Schr\"odinger equation. Here the vortex
wavefunction $\psi$ plays the part of its quantum counterpart, mass
density that of probability density, and the coordinate $\theta$ that
of the single valued phase.

One could object here, rightfully so, that the phase was derived through
the azimuthal flow field, whereas the Schr\"odinger equation emerges only after
suppressing $\theta_c$ and $\Phi_c$, leaving the kinematic plane-wave phase $\theta_d$ and the drift velocity potential $\Phi_d$.
Thus, the fluid Schr\"odinger equation has a restricted domain of validity: it
applies only in the regime where the drift velocity potential dominates over the local
azimuthal core velocity potential. Formally, at the level of velocity gradients, the far-field condition is
\begin{equation}
    |\nabla\Phi_c|\ll|\nabla\Phi_d|.
\end{equation}
Since
\begin{equation}
    |\nabla\Phi_c|=\frac{\ell}{r}=\frac{\hbar}{mr},
\end{equation}
and
\begin{equation}
    |\nabla\Phi_d|=v_d,
\end{equation}
this requires
\begin{equation}
    r\gg \frac{\hbar}{m v_d}
    =\frac{\lambda_{\rm db}}{2\pi}.
\end{equation}
Thus, for the specialized $J=\hbar$ Euler--Korteweg vortex, the vortex
Schr\"odinger equation is not a description of the near-core vortex
field itself, but a far-field approximation valid at distances large compared
with the reduced de Broglie wavelength scale associated with the vortex drift.
In this regime the core contribution still fixes the phase topology,
angular-momentum scale and capillary structure, but its local azimuthal velocity
field is subleading in the effective drift-wave equation.

The conventional Schr\"odinger equation does not possess an analogous near-field restriction in
its own formalism, but at the same time, it represents a structureless point particle. In conventional quantum theory, such internal or 'near-field'
structure, when relevant, is described by more developed frameworks rather than
by the Schr\"odinger equation itself.

\section{The Born rule}

The results in the previous section show a purely formal equivalence; the fluid Schr\"odinger is not a probabilistic equation
like its conventional quantum counterpart. Madelung’s original formulation employs
a probability density, whereas the classical Euler–Korteweg system
describes a mass density $\rho$. To see how the one and the other can be
related, consider the following construction.

Let $\rho_{v}(x,t)$ denote the mass density of a single vortex and
$\rho_{b}$ the background fluid density, with the vortex mass sharply
localized:
\begin{equation}
\rho_{v} \gg \rho_{b}
\label{eq:5.1}
\end{equation}

Consider some hypothetical analogue 'vortex double-slit experiment' where the vortex is sent through one of two possible openings, and its position is measured when it hits a screen behind the slits. The localization scale on the screen is large compared to the vortex core size, so we may model the vortex as a point mass:
\begin{equation}
\rho_{v} = m \, \delta(\zeta)
\label{eq:5.2}
\end{equation}

Where $\delta$ is the Dirac delta function, and:
\begin{equation}
\zeta = x - X(t)
\label{eq:5.3}
\end{equation}

Where $x$ is where the density is evaluated and $X(t)$ is the trajectory
followed by the vortex, with:
\begin{equation}
\dot{X} = v_{d}
\label{eq:5.4}
\end{equation}

So, given an initial vortex centre $\mathbf{X}_0$, the later centre $\mathbf{X}$ and density $\rho_v$ are uniquely determined. If we cannot determine $\mathbf{X}_0$ precisely, and moreover cannot determine both the vortex’s precise initial position and its drift velocity field (see section on the uncertainty principle), a hidden variable is introduced and with it an element of uncertainty.

Consider then a sequence of $n = 1,2,\dots,N$ trajectories with initial
positions $X_{0}^{n}$, trajectories
$X^{n}(t) = X(t ; X_{0}^{n})$. The distribution of mass density in the
limit $N \to \infty$ is:
\begin{equation}
\overline{\rho}_{v}(x,t)
=
\lim_{N\to\infty}
\sum_{n=1}^{N}
m\,\delta\!\left(x - X^{n}(t)\right)
\label{eq:5.5}
\end{equation}

And then the empirical distribution of positions of point vortices at
time $t$ is described as:
\begin{equation}
\mathcal{P}_{N}(x,t)
=
\frac{1}{N}
\sum_{n=1}^{N}
\delta\!\left(x - X^{n}(t)\right)
\label{eq:5.6}
\end{equation}

In the continuum limit this converges to a smooth density
$\mathcal{P}(x,t)$:
\begin{equation}
\mathcal{P} = \lim_{N\to\infty} \mathcal{P}_{N},
\qquad
\int \mathcal{P}(x,t)\,dx = 1
\label{eq:5.7}
\end{equation}

Where we see:
\begin{equation}
\mathcal{P} = \frac{1}{m}\,\overline{\rho}_{v}
\label{eq:5.9-pre}
\end{equation}

Thus $\mathcal{P}$ is the probability distribution of finding a single
vortex at a given position. Since $\overline{\rho}_{v}$ follows
the principle of local continuity, being a conserved quantity, and
$\mathcal{P}$ is separated from it only by a constant factor, probability
density must follow continuity as well. In this manner then, we can construct a continuity equation for $\mathcal{P}$, and insert $\mathcal{P}$ in our Euler--Korteweg equation instead of $\rho$.

Following the Madelung procedure:
\begin{equation}
\psi_{\mathcal{P}}
=
\sqrt{\mathcal{P}}\, e^{i\theta}
\begin{cases}
\displaystyle
\frac{\partial\Phi}{\partial t}
+ \frac{1}{2}(\nabla \Phi)^{2}
- \frac{\hbar^{2}}{2m^{2}}
  \frac{\nabla^{2}\sqrt{\mathcal{P}}}{\sqrt{\mathcal{P}}}
+ \frac{1}{m}V=0,
\\[1em]
\displaystyle
\frac{\partial \mathcal{P}}{\partial t}
+ \nabla\!\cdot(\mathcal{P}v) = 0
\end{cases}
\label{eq:5.9}
\end{equation}

The formal correspondence between the Schr\"odinger wavefunction and its fluid analogue is in this case mathematically exact, yielding the relation:
\begin{equation}
|\psi_{\mathcal{P}}|^{2} = \mathcal{P}
\label{eq:5.10}
\end{equation}

which is mathematically equivalent to the Born rule.\cite{born1926} In the present construction, however, $\mathcal P$ is a classical
ensemble density over repeated preparations with uncertain initial
conditions. Thus the Born rule for probability density appears here instead as a consequence of the
classical ensemble construction, following incomplete knowledge of initial conditions, not as a postulate driving true intrinsic quantum probability.

\section{Einstein--Planck relation}

Let us define the total specific energy:
\begin{equation}
\varepsilon
= \frac{1}{2}(\nabla \Phi)^{2}
+ h_{\rho}
- \frac{\hbar^{2}}{2m^{2}}
  \frac{\nabla^{2}\sqrt{\rho}}{\sqrt{\rho}}
+ V
\label{eq:6.1}
\end{equation}

Thus we have:
\begin{equation}
\frac{\partial \Phi}{\partial t}
= -\varepsilon
= -\frac{E}{m},
\qquad
m\,\frac{\partial \Phi}{\partial t}
= -m\varepsilon
= -E
\label{eq:6.2}
\end{equation}

Noting here we could define:
\begin{equation}
m\Phi = S
\label{eq:6.3}
\end{equation}

as the action of the system, \(S\), and \(E\) as \(H\), the real Hamiltonian, and the Hamilton–Jacobi equation of the vortex:
\begin{equation}
\frac{\partial S}{\partial t} = -H
\label{eq:6.4}
\end{equation}

\begin{equation}
H =
\frac{1}{2}\, m(\nabla \Phi)^{2}
+ m h_{\rho}
- \frac{\hbar^{2}}{2m}\,
  \frac{\nabla^{2}\sqrt{\rho}}{\sqrt{\rho}}
+ V
\label{eq:6.5}
\end{equation}

Taking the time derivative of the phase:
\begin{equation}
\theta = \frac{m}{\hbar}\Phi = \frac{S}{\hbar}
\quad\Rightarrow\quad
\frac{\partial \theta}{\partial t}
= \frac{m}{\hbar}\,\frac{\partial \Phi}{\partial t}
= \frac{1}{\hbar}\,\frac{\partial S}{\partial t}
\label{eq:6.6}
\end{equation}

We have:
\begin{equation}
\omega = \frac{v_{\theta}}{r}
\label{eq:6.7}
\end{equation}

And the material derivative of \(\theta\):
\begin{equation}
\frac{D\theta}{Dt}
= \frac{\partial \theta}{\partial t}
+ v_{r}\frac{\partial \theta}{\partial r}
+ v_{\theta}\frac{\partial \theta}{\partial \theta}
= \frac{\partial \theta}{\partial t}
+ \frac{v_{\theta}}{r}
\label{eq:6.8}
\end{equation}

For a steady vortex at rest, we have:
\begin{equation}
\frac{D\theta}{Dt} = 0,
\qquad
\frac{\partial \theta}{\partial t}
= -\frac{v_{\theta}}{r}
\label{eq:6.9}
\end{equation}

And thus we can express (rest) energy in terms of the vortex angular
frequency:
\begin{equation}
\omega = \frac{E}{\hbar},
\qquad
E = h f
\label{eq:6.10}
\end{equation}

Which matches the Einstein–Planck relation\cite{planck1900,einstein1905a}, relating angular frequency to total energy via \(\hbar\). Utilizing the radius \(r_{c} = \hbar/(mc)\) and inserting into eq.~\ref{eq:6.7} we can formulate:
\begin{equation}
\omega_{c}
= \frac{c}{r_{c}}
= \frac{m c^{2}}{\hbar}
\label{eq:6.12}
\end{equation}

And by equating these two expressions for the 'rest' angular frequency we obtain:
\begin{equation}
E = m c^{2}
\label{eq:6.13}
\end{equation}

Like Einstein’s famous formulation relating the rest mass of a body to its energy content\cite{einstein1905b}. These relations are direct consequences of the previously
stated assumptions and parameterizations of the model.
Interestingly, multiplying the isothermal equation of state by a volume \(u\):
\begin{equation}
P u = \rho u c^{2} = E = m c^{2}
\label{eq:6.11}
\end{equation}
Also shows that the total pressure energy contained in a volume of the fluid
equals \(mc^{2}\). Taking these as the rest frequency and wavelength of the vortex:
\begin{equation}
\omega_c = \frac{m c^{2}}{\hbar},
\qquad
\lambda_c = \frac{h}{m c}
\label{eq:7.9}
\end{equation}

We should note that it seems we cannot allow a spatial component to the rest wave to have:
\begin{equation}
\theta_c = \frac{m}{\hbar}\Phi_c = -\omega_c t
\quad\Rightarrow\quad
m\,\frac{\partial \Phi_c}{\partial t} = m c^{2}
\label{eq:7.11}
\end{equation}

Hence it seems that the rest-case core wave should be a standing wave, only oscillating in time. Possible solutions could be a torsional breathing mode instead of a radial one.  
Although this is an interesting inquiry, we will not investigate this further in this paper.

Finally we can note that for a counter-rotating vortex, the velocity, angular frequency and action change sign, and consequently the energy associated with the counter-rotating vortex becomes negative.

\section{De Broglie wavelength}
In what follows, the terms “rest” and “kinetic” are used in a fluid-mechanical sense: Rest refers to the frame comoving with the background fluid, while kinetic refers to the frame in which the background fluid has a drift velocity \(v_d\). Investigating the kinetic, drift wave field perturbations, we can see that
\begin{equation}
A e^{i\theta_d} = A e^{i(k_k x - \omega_k t)}
\label{eq:7.1}
\end{equation}

Takes the form of a plane wave ansatz, where
\begin{equation}
k_k = \frac{m v_d}{\hbar},
\qquad
\omega_k = \frac{1}{2} \frac{m v_d^{2}}{\hbar}
= \frac{\hbar k_k^{2}}{2m}
\label{eq:7.2}
\end{equation}

Where \(k_k\) and \(\omega_k\) are the ‘kinetic’ wavenumber and ‘kinetic’ angular frequency, and  
\(m v_d\) can be described as the momentum of the vortex \(p\), for the observer at rest with the far field fluid.  
And so the momentum and kinetic energy are described as:
\begin{equation}
p = \hbar k_k,
\qquad
E_k = \hbar \omega_k = \frac{p^{2}}{2m}
\label{eq:7.3}
\end{equation}

Analogous to the momentum and energy eigenvalues of QM’s Schr\"odinger equation.  
So that:
\begin{equation}
E = E_{\text{rest}} + E_k,
\qquad
\omega = \frac{m c^{2}}{\hbar} + \frac{1}{2}\frac{m v_d^{2}}{\hbar}
\label{eq:7.4}
\end{equation}

If we compare the difference in kinetic phase between two points separated by \(\Delta x\) we have:
\begin{equation}
\Delta \theta_d = k_k \Delta x
\label{eq:7.5}
\end{equation}

So for the distance between maxima modulo \(2\pi\):
\begin{equation}
2\pi = \frac{m v_d}{\hbar} \Delta x
\quad \Rightarrow \quad
\Delta x = \frac{2\pi \hbar}{m v_d}
\label{eq:7.6}
\end{equation}

We find the wavelength \(\Delta x\) is the de Broglie wavelength:
\begin{equation}
\lambda_{\mathrm{db}} = \frac{h}{m v_d}
\label{eq:7.7}
\end{equation}

With linear acoustic perturbations satisfying:
\begin{equation}
\omega = c k,
\qquad
k = \frac{2\pi}{\lambda}
\label{eq:7.8}
\end{equation}

The same form as lightwaves, which could be seen as analogous to the acoustic perturbations of the hypothetical fluid.  

\section{The uncertainty principle}

Heisenberg’s uncertainty principle\cite{heisenberg1927} is one of the fundamental
concepts in quantum mechanics, but it can also be applied to classical systems.
Let us construct a kinetic wave-packet of sorts, by superimposing many plane
wave ansatzes around a given wavenumber, to construct a sharply defined
‘particle’. Note that while the underlying near-core vortex dynamics remain non-linear,
the principle of linear superposition is applied here only to the drift waves
associated with the moving vortex. In this sense, the wave packet does not
construct the vortex core itself, but localizes a particle-like excitation
of the vortex drift-waves.

Taking the kinetic wavefunction:
\begin{equation}
\psi = \sqrt{\rho}\, e^{i\theta_d}
\label{eq:8.1}
\end{equation}

At \(t = 0\) we expand in wavenumber \(k = mv/\hbar\):
\begin{equation}
\psi(x,0) = \frac{1}{\sqrt{2\pi}} \int \phi(k)\, e^{ikx}\, dk
\label{eq:8.2}
\end{equation}

Where \(\phi(k)\) is the Fourier transform of \(\psi\):
\begin{equation}
\phi(k) = \frac{1}{\sqrt{2\pi}} \int \psi(x,0)\, e^{-ikx}\, dx
\label{eq:8.3}
\end{equation}

Defining the packet’s width in space \(\Delta x\) and wavenumber \(\Delta k\) by:
\begin{equation}
\Delta x^{2} = \int (x - X)^{2} |\psi(x,t)|^{2} dx,
\qquad
\Delta k^{2} = \int (k - k_{0})^{2} |\phi(k)|^{2} dk
\label{eq:8.4}
\end{equation}

Independently of any dynamics, the Cauchy–Schwarz (or Robertson–Schr\"odinger)
\cite{robertson1929, kennard1927, schrodinger1930}
inequality for a function and its Fourier transform gives the
Cramér–Rao bound \cite{cramer1946, hardy1933}:
\begin{equation}
\Delta x\, \Delta k \ge \frac{1}{2}
\label{eq:8.5}
\end{equation}

Following the parameterization:
\begin{equation}
\Delta p = \hbar\, \Delta k
\label{eq:8.6}
\end{equation}

So we can retrieve the classical analogue of the uncertainty principle:
\begin{equation}
\Delta x\, \Delta p \ge \frac{\hbar}{2}
\label{eq:8.7}
\end{equation}

Thus, the more we localize the packet in space (smaller \(\Delta x\)), the larger
\(\Delta p\).  
In the classical fluid picture, this is a purely mathematical consequence of
modeling the vortex as a particle-like superposition of waves via Fourier
transform \cite{ricaud2013}. 

\section{Lorentz transformation and Klein–Gordon}

The equations we derived in previous sections can be noted to be incomplete because they neglect the retardation of the change of the scalar fields associated with a vortex in drift. It is purely the physical demand to account for this retardation that allows us to formulate the Lorentz transformation.

Let us consider the vortex as a source S of an acoustic wave in the non-convective case:
\begin{equation}
\frac{\partial^2 \psi}{\partial t^2} - c^2 \nabla^2 \psi = S
\label{eq:11.1}
\end{equation}

Where \( S = q\,\delta(x) \), and \( q \) is the source strength. 
For the same perturbation, but in the fluid flowing with velocity \( v_{d} \):
\begin{equation}
\left( \frac{\partial}{\partial t} + v_d \frac{\partial}{\partial x} \right)^2 \psi - c^2 \nabla^2 \psi = S_d
\label{eq:11.2}
\end{equation}

Treating the core as a localized emitter with rest frequency \(\omega{_c} = mc^2 / \hbar\):
\begin{equation}
S_d(x,t) = q \delta(x - X) e^{-i \omega_c t}
\label{eq:11.3}
\end{equation}

Note that while $\omega_c$ follows from the parameterization of the model, interpreting
this frequency as an actual oscillatory mode is an additional assumption. In
the present work the mechanical nature of such a mode, if present, is left unspecified. However, vortices with a non-zero effective mass inherently exhibit intrinsic oscillatory dynamics, such as small-amplitude transverse oscillations.\cite{Richaud2021,Richaud2025,PhysRevE.111.034216} Thus, treating a massive vortex core as a localized oscillator is an assumption which has some physical grounds. 

Suppose we have the fluid in drift relative to the vortex source. In the frame attached to the fluid, the vortex is seen to be in motion along its x-axis, so that at the time t the vortex is at \(X = v_d t\), \(y = z = 0\). Its position at the retarded time \(\tau\) is:
\begin{equation}
\tau = t - r'/c
\label{eq:11.4}
\end{equation}

Where \(r'\) is the distance between the vortex and an observer stationary in (x, y, z) at the retarded time:
\begin{equation}
r' = \sqrt{(x - v_d \tau)^2 + y^2 + z^2}
\label{eq:11.5}
\end{equation}

And so:
\begin{equation}
c^2 (t - \tau)^2 = (x - v_d \tau)^2 + y^2 + z^2
\label{eq:11.6}
\end{equation}

Expanding, and collecting terms:
\begin{equation}
(v_d^2 - c^2)\tau^2 - 2(x v_d - c^2 t)\tau + x^2 + y^2 + z^2 - (ct)^2 = 0
\label{eq:11.7}
\end{equation}

Solving for \(\tau\):
\begin{equation}
(1 - M^2)\tau = t - \frac{M}{c} x - \frac{1}{c} \sqrt{(x - v_d t)^2 + (1 - M^2)(y^2 + z^2)}, \quad M = v_d/c
\label{eq:11.8}
\end{equation}

Where M is the Mach number of the fluid flow. So, with drift velocity constant, if the vortex centre were at rest with the fluid the scalar field \(\psi\) would be:
\begin{equation}
\psi = \frac{q}{4\pi} \frac{1}{\sqrt{x^2 + y^2 + z^2}}
\label{eq:11.9}
\end{equation}

In the convective case, it would become:
\begin{equation}
\psi = \frac{q}{4\pi} \frac{1}{\sqrt{(x - v_d t)^2 + (1 - M^2)(y^2 + z^2)}}
\label{eq:11.10}
\end{equation}

And thus the correct transformation between these frames, preserving the wave equation and accounting for the retarded field cannot be Galilean, but must be the Prandtl-Glauert-Lorentz transformation\cite{lorentz1904,poincare1905,einstein1905c,glauert1928,rienstra2025,feynmanlectures1964}:
\begin{equation}
x' = \gamma(x - v_d t), \quad y' = y, \quad z' = z, \quad t' = \gamma(t - v_d/c^2 \, x)
\label{eq:11.11}
\end{equation}

\begin{equation}
\gamma = \frac{1}{\sqrt{1 - M^2}}
\label{eq:11.12}
\end{equation}

If we utilize these coordinates to describe the phase:
\begin{equation}
\theta' = -\omega_c t' = -\omega_c \gamma(t - v_d/c^2 \, x) = \omega_c \gamma v_d/c^2 \, x - \gamma \omega_c t
\label{eq:11.13}
\end{equation}

Then if we probe the field at constant \(x = 0\) we find the boosted angular frequency:
\begin{equation}
\theta' = -\gamma \omega_c t, \quad \omega' = \gamma \omega_c
\label{eq:11.14}
\end{equation}

Since in this case the introduced spatial component, the time derivative of the kinetic wavenumber \(k'\) is zero:
\begin{equation}
k' = \gamma (m v_d)/\hbar
\label{eq:11.15}
\end{equation}

And since in this transformed frame the fluid Planck - Einstein relation holds, we also find the fluid ‘relativistic’ energy, mass and momentum:
\begin{equation}
E' = \gamma m c^2, \quad m' = \gamma m, \quad p' = \gamma m v_d
\label{eq:11.16}
\end{equation}

Furthermore we can in a usual manner start from:
\begin{equation}
m' = \frac{m}{\sqrt{1 - (v_d^2)/c^2}}
\label{eq:11.17}
\end{equation}

\begin{equation}
{m'}^2 = \frac{m^2}{1 - (v_d^2)/c^2} \;\rightarrow\; {m'}^2 - \frac{{m'}^2 v_d^2}{c^2} = m^2
\label{eq:11.18}
\end{equation}

And, multiplying by \(c^4\) and rearranging find the energy-momentum relation:
\begin{equation}
{m'}^2 c^4 - {m'}^2 v_d^2 c^2 = m^2 c^4 \;\rightarrow\; {E'}^2 = m^2 c^4 + {p'}^2 c^2
\label{eq:11.19}
\end{equation}

And the dispersion relation:
\begin{equation}
{\omega'}^2 = c^2 {k'}^2 + \omega_c^2
\label{eq:11.20}
\end{equation}

Which are the Klein–Gordon equations\cite{klein1926,gordon1926} in momentum space. To infer the coordinate-space KG equation, note that a ‘monochromatic’ solution:
\begin{equation}
\psi' = e^{-i\omega' t' + ik' x'}
\label{eq:11.21}
\end{equation}

Satisfies:
\begin{equation}
\frac{\partial^2 \psi'}{\partial t'^2} - c^2 \frac{\partial^2 \psi'}{\partial x'^2} = -\omega_c^2 \psi'
\label{eq:11.22}
\end{equation}

Note that this should not be read as a derivation of Klein--Gordon field theory,
which would require a field Lagrangian or Hamiltonian formulation, quantization in terms of field operators and creation and annihilation modes.

Following the above we could say that at fixed x in the lab frame the frequency of oscillation increases by the Lorentz factor compared to the rest phase frequency. It is because the wave field changes in the convective case that the observing frame perceives an increase in energy, mass and momentum.

As in relativity, we can note that if we instead investigate the ticks of the oscillating core ‘clock’ along the vortex ‘worldline’, \(X = v_d t\):
\begin{equation}
t' = \gamma(t - v_d/c^2 \, X) = \gamma(1 - (v_d^2)/c^2)t = t/\gamma
\label{eq:11.23}
\end{equation}

And so evaluated at this translational X we instead have,
\begin{equation}
\theta_{(x')}^{'} = -\omega_c t' = -\omega_c \, t/\gamma, \quad \frac{d\theta_{(x')}^{'}}{dt} = -\omega_c/\gamma
\label{eq:11.24}
\end{equation}

That is we could say due to the transform, we should hold the 'clock' riding along the vortex core to be dilated by the Lorentz factor in relation to the 'clock' at rest with the fluid. And in much the same manner we could also obtain an analogue relativistic length contraction.

And we can see, by Taylor expanding the boosted angular frequency, neglecting higher order terms:
\begin{equation}
\omega' - \omega_c = \gamma \omega_c - \omega_c = \omega_c (1 + \tfrac{1}{2} M^2 + \cdots) - \omega_c \approx \tfrac{1}{2} (m v_d^2)/\hbar
\label{eq:11.25}
\end{equation}

That the guiding plane waves that came into the Schr\"odinger equations are simply the low velocity approximation of the boosted rest wave field. Seemingly, the standing wave core is a source of propagating waves when excited by a drift velocity.

%%%%%%%%%%%%%%%%%%%%%%%%%%%%%%%%%%%%%%%%%%
\section{Discussion}

The results of this paper show that some of the entities and rules associated with quantum theory and relativity have a mathematically consistent set of analogies in a classical continuum model. The Schr\"odinger equation is in this model a convenient linear combination of two more fundamental fluid equations: the continuity and momentum equations of a capillary vortex. We can also clearly see its limitations. The equation neglects the `rest' rotational velocity field that is essential to its own derivation. It is only a weak-field approximation, breaking down at high Mach numbers where it is a mathematical necessity to employ a Lorentz transform, when transforming between two frames with different relative velocities to the fluid.

This paper is proof of nothing other than a mathematical equivalence between a very specific fluid model and Schr\"odinger’s and Klein–Gordon’s equations, but might invite speculation regarding physical interpretation. Could we possibly provide an explanation for the quantum behavior of elementary particles and the relativistic dilation of clocks, if we infer that our `elementary' particles are these vortices – excitations of a more fundamental, material continuum? It is very improbable though, that such a physical interpretation would survive closer scrutiny.

First of all, an exact correspondence between the conventional Schr\"odinger/KG and the fluid equations hinges on the implicit assumption that the laboratory where vortices are seen to be in motion with $v_d$, is at rest with the fluid. In this frame, the acoustic Lorentzian spacetime derived by Unruh and Visser is not only experienced by acoustic perturbations: If source and receiver, ruler, clock and observer are all (constellations of) these vortices, subject to the Lorentzian acoustic metric, then this experience should be ours as well. But then the rest frame of the fluid model is a preferred frame, which puts very stringent bounds on the compatibility of a physical interpretation of the fluid model and tests of relativity.

Furthermore Schr\"odinger’s and Klein–Gordon are perhaps the simplest formulations of quantum mechanics, published precisely a 100 years ago. Since then theory has developed and expanded significantly. A physical interpretation would require extending the analogy to recover to a major extent the Standard Model, with analogues for spin, antimatter and CP-symmetry, the Pauli exclusion principle, tunneling, entanglement and Cooper pairing just a few of the many challenges that it would face.

It should also be noted that while a fluid interpretation and Bohmian mechanics seem much aligned, both causal and with hidden variables, both supposing that the pilot wave of one particle or velocity field of one vortex depends on the position of all others, they differ in a crucial sense: Describing a multi-body field in configuration space, as is conventional in Bohmian/quantum mechanics\cite{Goldstein2025Bohmian}, would be physically incomplete and generally incorrect for the fluid model. Without a formulation in configuration space, even with the vortices all being excitations of the same global field, and Korteweg stress arising from a non-local kernel interaction\cite{Giovangigli2020}, the theory should be Bell-local: Any change of field over some distance should be constrained by the speed of sound. Experimental tests of Bell inequalities\cite{bell1964} place extremely strong empirical constraints on any local hidden-variable models.

More importantly, beyond questions of interpretation, formal analogies have historically played a productive role in theoretical physics. The identification of  correspondences between distinct fields of physics can, in certain cases, allow the transfer of mathematical or pedagogical insight across otherwise disparate domains. In the present case, casting idealized vortex equations into a form mathematically equivalent to Schr\"odinger’s and Klein–Gordon’s equations may offer a complementary perspective on both sides of the analogy.

%%%%%%%%%%%%%%%%%%%%%%%%%%%%%%%%%%%%%%%%%%
\section{Conclusions}

In conclusion, the momentum and continuity equations governing an Euler--Korteweg vortex can be written in a Schr\"odinger-type form through a reverse Madelung transform. This result is obtained by considering an irrotational vortex in an inviscid, barotropic, isothermal fluid, where a steep pressure gradient at the core causes Korteweg stress, with the vortex being excited by a drift velocity. For arbitrary constant angular momentum $J$ and sound speed $c_s$, this gives a general class of Schr\"odinger-type equations. Under the parameterization
$J=\hbar$ and $c_s=c$, together with a ratio of characteristic lengths giving a capillary coefficient 
$\kappa=\hbar^2/(4m^2)$, this correspondence becomes formally identical to the
Schr\"odinger equation in the far-field approximation.

Following incomplete knowledge of the initial conditions over an ensemble of vortices, the same reverse Madelung transform can be used to yield the exact probability-density form of the Schr\"odinger equation, with an analogue relation
$|\psi_{\mathcal P}|^2=\mathcal P$ that is mathematically equivalent to the Born rule. Hydrodynamic identities analogous to well-known identities such as the
Einstein--Planck relation, $E=mc^2$, Bohm's guiding equation and the
de Broglie wavelength all emerge from this model, given the chosen
parameterization. If one superimposes the linear drift waves associated with the vortex into a particle-like wave packet, one obtains a classical analogue of the uncertainty principle.

Furthermore, if one assumes that the characteristic frequency
$\omega_c=mc^2/\hbar$ may be associated with a localized rest-phase oscillation
of the vortex, then accounting for retardation of the vortex wave field in uniform convection
requires the Lorentz transformation and yields the Klein--Gordon equations, with the Schr\"odinger equation appearing as the low-Mach-number limit approximation.

These results demonstrate that there is an interesting formal correspondence between vortices and quantum wave equations, sufficiently structured to warrant further inquiry. It seems of great interest to further investigate the analogy, to see if it may provide additional (pedagogical) insights into analogous behavior between classical wave systems, superfluid vortex dynamics, topology, quantum mechanics and relativity.


%\bibliographystyle{ptephy}
%\bibliography{bibliography}
%
% once the .bbl file has been generated then place the text in your article.

\vspace{0.2cm}
\noindent

\let\doi\relax

%This is added by T. Yoneya (editor-in-chief) on 2020/07/09.

%without this code before the command "\begin{thebibliography}{}" , an error will be %flagged. When the bibliography is provided as separate .bib file, then this code %should be placed above the commands "\bibliographystyle{}" and "\bibliography{}" %inside the main TeX file. 

\begin{thebibliography}{10}

\bibitem{barcelo2005}
C.~Barcel{\'o}, S.~Liberati, and M.~Visser, Living Reviews in Relativity, {\bf 8}, 12 (2005).

\bibitem{unruh1976}
W.~G. Unruh, Physical Review D, {\bf 14}, 870--892 (1976).

\bibitem{unruh1981}
W.~G. Unruh, Physical Review Letters, {\bf 46}, 1351--1353 (1981).

\bibitem{unruh1995}
W.~G. Unruh, Physical Review D, {\bf 51}, 2827--2838 (1995).

\bibitem{visser1993}
M.~Visser, arXiv preprint (1993),  {{gr-qc/9311028}}.

\bibitem{visser1998}
M.~Visser, Classical and Quantum Gravity, {\bf 15}, 1767--1791 (1998).

\bibitem{weinfurtner2011}
S.~Weinfurtner, E.~W. Tedford, M.~C.~J. Penrice, W.~G. Unruh, and G.~A. Lawrence, Physical Review Letters, {\bf 106}, 021302 (2011).

\bibitem{jacobson1995}
T.~Jacobson, Physical Review Letters, {\bf 75}, 1260--1263 (1995).

\bibitem{debroglie1923}
L.~de~Broglie, Comptes Rendus, {\bf 177}, 507--510 (1923).

\bibitem{schrodinger1926a}
E.~Schr{\"o}dinger, Annalen der Physik, {\bf 384}, 361--376 (1926).

\bibitem{schrodinger1926b}
E.~Schr{\"o}dinger, Annalen der Physik, {\bf 384}, 489--527 (1926).

\bibitem{madelung1927}
E.~Madelung, Zeitschrift f{\"u}r Physik, {\bf 40}, 322--326 (1927).

\bibitem{takabayasi1952}
T.~Takabayasi, Progress of Theoretical Physics, {\bf 8}, 143--182 (1952).

\bibitem{bohm1952a}
D.~Bohm, Physical Review, {\bf 85}, 166--179 (1952).

\bibitem{bohm1952b}
D.~Bohm, Physical Review, {\bf 85}, 180--193 (1952).

\bibitem{couder2005}
Y.~Couder, S.~Proti{\`e}re, E.~Fort, and A.~Boudaoud, Nature, {\bf 437}, 208 (2005).

\bibitem{couder2006}
Y.~Couder and E.~Fort, Physical Review Letters, {\bf 97}, 154101 (2006).

\bibitem{bush2015}
J.~W.~M. Bush, Annual Review of Fluid Mechanics, {\bf 47}, 269--292 (2015).

\bibitem{onsager1949}
L.~Onsager, Il Nuovo Cimento, {\bf 6}, 279--287 (1949).

\bibitem{feynman1955}
R.~P. Feynman,
\newblock Application of quantum mechanics to liquid helium,
\newblock In C.~J. Gorter, editor, {\em Progress in Low Temperature Physics}, volume~1, pages 17--53. North-Holland, Amsterdam (1955).

\bibitem{fetter2001}
A.~L. Fetter and A.~A. Svidzinsky, Journal of Physics: Condensed Matter, {\bf 13}, R135--R194 (2001).

\bibitem{popov1973}
V.~N. Popov, Zh. Eksp. Teor. Fiz., {\bf 64}, 672--680, English transl.: Sov. Phys. JETP 37, 341--345 (1973).

\bibitem{benzoni2005}
S.~Benzoni-Gavage, S.~Descombes, D.~Jamet, and L.~Mazet, Interfaces and Free Boundaries, {\bf 7}, 371--414 (2005).

\bibitem{carles2012}
R.~Carles, R.~Danchin, and J.-C. Saut, Nonlinearity, {\bf 25}, 2843--2873 (2012).

\bibitem{bresch2019}
D.~Bresch, M.~Gisclon, and I.~Lacroix-Violet, Archive for Rational Mechanics and Analysis, {\bf 233}, 975--1025 (2019).

\bibitem{mauri2021}
R.~Mauri, Foundations of Physics, {\bf 51}, 81 (2021).

\bibitem{gui2025}
G.~Gui and T.~Tang, Journal of Differential Equations, {\bf 2025}, 330--347 (2025).

\bibitem{vanderwaals1893}
J.~D. van~der Waals, Zeitschrift f{\"u}r Physikalische Chemie, {\bf 13}, 657--725 (1893).

\bibitem{korteweg1901}
D.~J. Korteweg, Archives N{\'e}erlandaises des Sciences Exactes et Naturelles, {\bf 6}, 1--24 (1901).

\bibitem{dunnserrin1985}
J.~E. Dunn and J.~Serrin, Archive for Rational Mechanics and Analysis, {\bf 88}, 95--133 (1985).

\bibitem{10.1063/5.0271672}
P.~Vadasz, Physics of Fluids, {\bf 37}(8), 086123 (08 2025).

\bibitem{thouless2007}
D.~J. Thouless, Physical Review Letters, {\bf 99}, 105301 (2007).

\bibitem{Richaud2020}
A.~Richaud, V.~Penna, R.~Mayol, and M.~Guilleumas, Phys. Rev. A, {\bf 101}, 013630 (Jan 2020).

\bibitem{Richaud2021}
A.~Richaud, V.~Penna, and A.~Fetter, Physical Review A, {\bf 103}(2), 023311 (2021).

\bibitem{kanjo2024}
A.~Kanjo and H.~Takeuchi, Physical Review A, {\bf 110}, 063311 (2024).

\bibitem{levrouw2025}
L.~Levrouw, H.~Takeuchi, and J.~Tempere, Physical Review A, forthcoming (2025),  {{2505.12590}}.

\bibitem{Richaud2025}
Andrea Richaud, Matteo Caldara, Massimo Capone, Pietro Massignan, and Gabriel Wlazłowski, Physical Review A, {\bf 112}, L051306 (2025).

\bibitem{PhysRevE.111.034216}
J.~D'Ambroise, W.~Wang, C.~Ticknor, R.~Carretero-Gonz{\'a}lez, and P.~G. Kevrekidis, Phys. Rev. E, {\bf 111}, 034216 (2025).

\bibitem{cahn1958}
J.~W. Cahn and J.~E. Hilliard, Journal of Chemical Physics, {\bf 28}, 258--267 (1958).

\bibitem{cahn1959}
J.~W. Cahn and J.~E. Hilliard, Journal of Chemical Physics, {\bf 30}, 1121--1125 (1959).

\bibitem{rowlinson1982}
J.~S. Rowlinson and B.~Widom,
\newblock {\em Molecular Theory of Capillarity},
\newblock  (Clarendon Press, Oxford, 1982).

\bibitem{sengers1991}
J.~V. Sengers, J.~M.~J. van Leeuwen, and J.~W. Schmidt, Physica A, {\bf 172}, 20--29 (1991).

\bibitem{chaikin1995}
P.~M. Chaikin and T.~C. Lubensky,
\newblock {\em Principles of Condensed Matter Physics},
\newblock  (Cambridge University Press, Cambridge, 1995).

\bibitem{onuki2002}
A.~Onuki,
\newblock {\em Phase Transition Dynamics},
\newblock  (Cambridge University Press, Cambridge, 2002).

\bibitem{born1926}
M.~Born, Zeitschrift f{\"u}r Physik, {\bf 37}, 863--867 (1926).

\bibitem{planck1900}
M.~Planck, Verhandlungen der Deutschen Physikalischen Gesellschaft, {\bf 2}, 237--245 (1900).

\bibitem{einstein1905a}
A.~Einstein, Annalen der Physik, {\bf 17}, 132--148 (1905).

\bibitem{einstein1905b}
A.~Einstein, Annalen der Physik, {\bf 18}, 639--641 (1905).

\bibitem{heisenberg1927}
W.~Heisenberg, Zeitschrift f{\"u}r Physik, {\bf 43}, 172--198 (1927).

\bibitem{robertson1929}
H.~P. Robertson, Physical Review, {\bf 34}, 163--164 (1929).

\bibitem{kennard1927}
E.~H. Kennard, Zeitschrift f{\"u}r Physik, {\bf 44}, 326--352 (1927).

\bibitem{schrodinger1930}
E.~Schr{\"o}dinger, Sitzungsberichte der Preussischen Akademie der Wissenschaften, Physikalisch-Mathematische Klasse, pages 296--303 (1930).

\bibitem{cramer1946}
H.~Cram{\'e}r,
\newblock {\em Mathematical Methods of Statistics},
\newblock  (Princeton University Press, Princeton, 1946).

\bibitem{hardy1933}
G.~H. Hardy, Journal of the London Mathematical Society, {\bf 8}, 227 (1933).

\bibitem{ricaud2013}
B.~Ricaud and B.~Torr{\'e}sani, Advances in Computational Mathematics, {\bf 39}, 431--485 (2013).

\bibitem{lorentz1904}
H.~A. Lorentz, KNAW Proceedings, {\bf 6}, 809--831 (1904).

\bibitem{poincare1905}
H.~Poincar{\'e}, Comptes Rendus, {\bf 140}, 1504--1508 (1905).

\bibitem{einstein1905c}
A.~Einstein, Annalen der Physik, {\bf 17}, 891--921 (1905).

\bibitem{glauert1928}
H.~Glauert, Proceedings of the Royal Society A, {\bf 118}, 113--119 (1928).

\bibitem{rienstra2025}
S.~W. Rienstra and A.~Hirschberg,
\newblock {\em An Introduction to Acoustics},
\newblock  (Eindhoven University of Technology, Eindhoven, 2025).

\bibitem{feynmanlectures1964}
R.~P. Feynman, R.~B. Leighton, and M.~Sands,
\newblock {\em The Feynman Lectures on Physics, Vol. II},
\newblock  (Addison-Wesley, Reading, 1964).

\bibitem{klein1926}
O.~Klein, Zeitschrift f{\"u}r Physik, {\bf 37}, 895--906 (1926).

\bibitem{gordon1926}
W.~Gordon, Zeitschrift f{\"u}r Physik, {\bf 40}, 117--133 (1926).

\bibitem{Goldstein2025Bohmian}
S.~Goldstein,
\newblock Bohmian mechanics (2025),
\newblock Fall 2025 Edition, The Stanford Encyclopedia of Philosophy.

\bibitem{Giovangigli2020}
V.~Giovangigli, Physical Review E, {\bf 102}(1), 012110 (2020).

\bibitem{bell1964}
J.~S. Bell, Physics, {\bf 1}, 195--200 (1964).

\end{thebibliography}

\end{document}